\documentclass[aps,amsmath,pra,nofootinbib,reprint,longbibliography]{revtex4-2}
\usepackage{graphicx}
\usepackage{dcolumn}
\usepackage{bm}
\usepackage{amssymb} 
\usepackage{verbatim}
\usepackage{dsfont}
\usepackage{wrapfig}
\usepackage{epstopdf,nicefrac}
\usepackage{lipsum}
\usepackage{epstopdf}
\usepackage{natbib}
\usepackage{lipsum}
\RequirePackage{float}
\usepackage[colorlinks=true,urlcolor=blue,linkcolor=red,citecolor=blue]{hyperref}

\begin{document}
	\title{Quantum coherence measures in entangled atomic systems}
	
	\author{Arnab Mukherjee}
	\email{arnab.mukherjee@bose.res.in}
	
	\author{Soham Sen}
	\email{sensohomhary@gmail.com}
	
	\author{Sunandan Gangopadhyay}
	\email{sunandan.gangopadhyay@gmail.com}
	
	\affiliation{
		Department of Astrophysics and High Energy Physics\\
		S. N. Bose National Centre for Basic Sciences, JD Block, Sector-III, Salt Lake, Kolkata 700106, India} 

\begin{abstract}
    \noindent In this study, we investigate the effect of the Lorentz transformation on the measures of quantum coherence in an entangled atomic system. Here, we consider the effect of  this relativistic boosts on two-particle entangled generalized Gaussian wave packets in two scenarios. In the first scenario, we consider that the relativistic boost affects the one particle and other remains unaffected while in the second scenario, we consider that both the particles are affected by the effect of the relativistic boost. The coherence of the wave function as measured by the boosted observer is studied as a function of the boost parameter and the width of the Gaussian wave packets. Using various formulations of coherence, it is shown that in general the coherence decays with increase in the width of the Gaussian wave packet, higher values of boost parameter, and the number of particles on which boost is applied.  
\end{abstract}

\maketitle

\section{Introduction}
\noindent Spanning a century of groundbreaking breakthroughs, quantum entanglement has emerged as the cornerstone of contemporary quantum information science and technology. Quantum information science, as we know it today, started from a question asked by Albert Einstein. He thought that there may be local hidden variables in nature that resolves the indeterministic nature of quantum mechanics. In their 1935 paper \cite{EPR}, Einstein, Podolsky, and Rosen argued that quantum mechanics is not complete. At the same time, the paper regarding entanglement between two separated states had been investigated in \cite{Schrodinger_1935}. In the seminal 1964 paper by Bell \cite{BellEPR}, he obtained some constraint relations that arose as a consideration of hidden variables (local) in case of the calculation of the correlation between two quantum mechanical particles. It was further argued that no such hidden variables can exist \cite{BellEPRViolation}.  These works eventually led to the birth of quantum information theory. Through the advances of the quantum information science, it has been observed that quantum resources are indispensable for performing any operational task. Over the past two decades, studies spanning from quantum cryptography to quantum metrology, it has been proven that among many other quantum resources quantum entanglement and quantum coherence can play significant role as quantifiers. In recent few years, analyses involving quantum entanglement in a relativistic setting have gained significant momentum 
\cite{EPRBExperiment,GingrichAdami,PeresScudoTerno,
JordanShajiSudarshan,AhnLeeMoonHwang,AhnLeeHwang,
AlsingMilburn,LeeYoung,Chakrabarti_2009,Czachor,CzachorWilczewski,
FuentesMann,AlsingFuentesMannTessier,CabanRembielinski,
CabanRembielinski2,JordanShajiSudarshan2,LamataMartinSolano,
MoonHwangAhn}. Even if quantum mechanics as a whole is a non-relativistic theory, quantum entanglement is dependent on the reference frame of the observer. Under Lorentz transformation, the momentum of the particle changes, and it was observed in \cite{GingrichAdami,PeresTerno} that due to this change in the momentum the spin of  a massive particle gets altered as well. In \cite{FuentesMann,AlsingFuentesMannTessier,AdessoFuentesEricsson}
it was thoroughly observed that entanglement between particles is dependent upon the observer. Another important aspect that one can investigate is the quantum coherence measure which can be used to execute several quantum tasks \cite{BrandaoGour,HorodeckiOppenHeim,GourSpekkens,
GourMarvianSpekkens,MarvianSpekkens,MarvianSpekkens2,
MarvianSpekkens3,MarvianSpekkensZanardi,MarvianSpekkens4}.
Among the known measurements of quantum coherence, it can either be basis-dependent \cite{ChitambarGour} or basis-independent\cite{YaoDongXiaoSun,WangFangYu}.

\noindent In a recent work \cite{MSG2024quantum}, we have considered spin-momentum entangle state for a single particle where the entire system is given a relativistic boost. Initially, in \cite{chatterjee2017preservation}, the single-particle state with a Gaussian wave packed under a relativistic boost has been considered. In \cite{MSG2024quantum}, we have investigated the single particle model when the states are given by generalized Gaussian wave packets. In our work, we observe that the value of the measured coherence by the boosted observer decays with the increasing value of the momentum as well as the value of the boost parameter. The generalized Gaussian wave-packet considered in our case has the analytical structure as $\mathcal{N}_n p^ne^{-\nicefrac{p^2}{\sigma^2}}$ with $\mathcal{N}_n$ denoting the normalization constant and $\sigma$ denoting the Gaussian width. For such a simple system, we find out that the $l_1$ norm measure of coherence is exactly equal to the Frobenius norm measure of coherence and observed that for system with the values of the diagonal elements of the reduced density matrix being equal to half, such coincidence can occur provided the off-diagonal elements are complex conjugate to each other. Using the properties of the coherence measures, we also obtain a bound on the generalization parameter in a (1+1)-dimensional as well as (3+1)-dimensional analysis. We observe for all the cases that with increase in the value of the generalization parameter $n$, the loss of coherence becomes even more significant. The general extension of our work \cite{MSG2024quantum}, is to extend the model system to a two-particle system. In \cite{FriisBertlmannHuber}, the two-particle model under Lorentz boost was considered and the behaviour of the Bell inequalities were investigated from the perspective of different observers. 

\noindent In this work, we adapt the two-particle entangled model system under the Lorentz boost where the boost is applied along a single and both of the directions corresponding to the momentum of the two particles. The coherence measures for the bipartite quantum state under the Lorentz boost scenario is then observed. 

\noindent The manuscript is organized as follows. In section (\ref{S2}), we review the case where a single particle state is under the effect of the Lorentz boost and discuss the basic formalism. In section (\ref{S3}), we analyze the case of Lorentz boost for a bipartite quantum state where we consider the case when a single particle is under Lorenz boost along with case when both the particles are under the effect of Lorentz boost. In section (\ref{S4}), we then discuss the effect of Lorentz boost on different quantum coherence measures. Finally, in section (\ref{S5}), we discuss our findings and then conclude in section (\ref{sec:Conclu}). 
\section{Single particle quantum state under Lorentz boost}\label{S2}
\noindent In this section, we are going to give a brief review about the situation when a single particle quantum state experiences the effect of Lorentz boost. See \cite{MSG2024quantum} for a detailed derivation.

Consider the single particle quantum state in the given form 
\begin{eqnarray}
\vert \Psi \rangle =\vert \mathbf{p}\rangle \otimes \vert \sigma \rangle \equiv \vert \mathbf{p},\sigma\rangle~.
\end{eqnarray}
Here, the spatial part of the four‐momentum \(p^{\mu}\) is written as \(\mathbf{p}\), its temporal component is  
\(p^0 = \sqrt{\mathbf{p}^2 + m^2}\,,\) and the spin of a massive particle is labelled by \(\sigma\). The states \(\lvert \mathbf{p},\sigma\rangle\) are eigenvectors of the four‐momentum operator \(P^{\mu}\) with eigenvalues \(p^{\mu}\), and they satisfy the relation  
\begin{eqnarray}
P^{\mu}\vert \mathbf{p},\sigma\rangle =p^{\mu}\vert \mathbf{p},\sigma\rangle\,.
\end{eqnarray}

\noindent Following the method given in \cite{weinberg1995quantum}, the single-particle quantum state $\vert \mathbf{p},\sigma \rangle$, under the Lorentz boosts $\Lambda$, takes the form
\begin{eqnarray}
\mathcal{U}(\Lambda)\vert \mathbf{p},\sigma \rangle=\sqrt{\frac{(\Lambda p)^0}{p^0}}\displaystyle\sum_{\sigma'}D_{\sigma'\sigma}(W(\Lambda,\mathbf{p}))\vert \Lambda \mathbf{p},\sigma' \rangle\,.\label{t-state}
\end{eqnarray}
After a Lorentz transformation \(\Lambda\), the spatial momentum becomes \(\Lambda\mathbf{p}\), and \(W(\Lambda,\mathbf{p})\) labels the corresponding element of the little‐group representation \(D\bigl(W(\Lambda,\mathbf{p})\bigr)\). For particles with nonzero mass, the little group is isomorphic to \(SO(3)\), so \(W(\Lambda,\mathbf{p}) \in SO(3),\)
while \(D\bigl(W(\Lambda,\mathbf{p})\bigr)\) is represented through the double cover \(SU(2)\).

\noindent From \cite{chatterjee2017preservation, MSG2024quantum}, we found that the unitary representation of the little group $D(W(\Lambda,\mathbf{p}))$ takes the form
\begin{eqnarray}\label{little_group}
D(W(\Lambda,\mathbf{p}))=\cos \frac{\phi}{2}\mathds{1}+i\sin\frac{\phi}{2}(\Sigma\cdot\hat{n})
\end{eqnarray}
where
\begin{eqnarray}\label{components}
\cos\frac{\phi}{2}&=&\frac{\cosh\frac{\alpha}{2}\cosh\frac{\chi}{2}+\sinh\frac{\alpha}{2}\sinh\frac{\chi}{2}(\hat{e}\cdot\hat{f})}{\sqrt{\frac{1}{2}+\frac{1}{2}\cosh\alpha\cosh\chi+\frac{1}{2}\sinh\alpha\sinh\chi(\hat{e}\cdot\hat{f})}}\nonumber\\
&\,&\\
\sin\frac{\phi}{2}\hat{n}&=&\frac{\sinh\frac{\alpha}{2}\sinh\frac{\chi}{2}(\hat{e}\times\hat{f})}{\sqrt{\frac{1}{2}+\frac{1}{2}\cosh\alpha\cosh\chi+\frac{1}{2}\sinh\alpha\sinh\chi(\hat{e}\cdot\hat{f})}}\nonumber\\
&\,&
\end{eqnarray}
with $\phi,\,\hat{n}$ being the angle and axis of the Wigner rotation respectively, when we consider the four-momentum of the particle in the form
\begin{eqnarray}
p^{\mu}=(m \cosh \chi, m \sinh \chi \hat{f})~.
\end{eqnarray}
Here, $m$ is the mass of the particle, and the velocity of the moving frame $\mathcal{O'}$, $\vec{v}=\tanh\alpha \,\hat{e}$.

\noindent To proceed further, we take a pure state in an inertial frame $\mathcal{O}$ as
\begin{eqnarray}
\vert \Psi\rangle 
=\displaystyle\sum_{\sigma}\int d^3\mathbf{p}\, \psi(\mathbf{p})c_{\sigma}\vert \mathbf{p},\sigma\rangle\,.\label{psi}
\end{eqnarray}
Using the above state, the density matrix can be calculated as
\begin{eqnarray}
\varrho&=&\displaystyle\sum_{\sigma_1,\sigma_2}\int d^3\mathbf{p}_{1}\int d^3\mathbf{p}_{2}\, \psi(\mathbf{p}_{1})\psi^{\ast}(\mathbf{p}_{2})c_{\sigma_1}c^{\ast}_{\sigma_2}\nonumber\\&\times&\vert \mathbf{p}_{1},\sigma_{1}\rangle\langle \mathbf{p}_{2},\sigma_{2}\vert\nonumber~.\\\label{varrho}
\end{eqnarray}
To calculate the reduced density matrix for the spin degrees of freedom, we take partial trace over the momentum degrees of freedom, the reduced density matrix then becomes
\begin{eqnarray}
\rho&=&\text{Tr}_p[\vert\Psi\rangle\langle\Psi\vert]=\int d^3\mathbf{p}\langle \mathbf{p}\vert\Psi\rangle\langle\Psi\vert\mathbf{p}\rangle\nonumber\\
&=&\displaystyle\sum_{\sigma_1,\sigma_2}\int d^3\mathbf{p}\,\bigl|\psi(\mathbf{p})\bigr|^{2}\;c_{\sigma_1}c^{\ast}_{\sigma_2}\vert\sigma_{1}\rangle\langle\sigma_{2}\vert\,.\label{rho}
\end{eqnarray}
Using eq.~\eqref{t-state}, the single particle quantum state 
with respect to a boosted reference frame $\mathcal{O'}$ can be represented as
\begin{eqnarray}
\vert \Psi_{\Lambda}\rangle&=&\displaystyle\sum_{\sigma}\int d^3\mathbf{p}\, \psi(\mathbf{p})\;c_{\sigma}\sqrt{\frac{(\Lambda p)^0}{p^0}}\displaystyle\nonumber\\&\otimes&\sum_{\sigma'}D_{\sigma'\sigma}(W(\Lambda,\mathbf{p}))\vert \Lambda \mathbf{p},\sigma' \rangle\,.\nonumber\\\label{boostedpsi}
\end{eqnarray}
Therefore, the density matrix in this boosted reference frame is given by
\begin{eqnarray}
\varrho_{\Lambda}&=&\displaystyle\sum_{\sigma_1,\sigma_2}\int d^3\mathbf{p}_{1}\int d^3\mathbf{p}_{2}\,\sqrt{\frac{(\Lambda p_1)^0(\Lambda p_2)^0}{p_1^0p_2^0}} \psi(\mathbf{p}_{1})\psi^{\ast}(\mathbf{p}_{2})\nonumber\\
&\times & c_{\sigma_1}c^{\ast}_{\sigma_2}\displaystyle\sum_{\sigma'_1,\sigma'_2}\;D_{\sigma'_1\sigma_1}(W(\Lambda,\mathbf{p}_1))\vert \Lambda\mathbf{p}_{1},\sigma'_{1}\rangle\langle \Lambda\mathbf{p}_{2},\sigma'_{2}\vert\nonumber\\
&\times&\;D_{\sigma'_2\sigma_2}^{\dagger} (W(\Lambda,\mathbf{p}_2))~.\label{boostedvarrho}
\end{eqnarray} 
By tracing out the momentum degrees of freedom, and using the relation 
\begin{eqnarray}
\delta^{(3)}(\Lambda\mathbf{p}_{1}-\Lambda\mathbf{p}_{2})=\frac{p_1^0}{(\Lambda p_1)^0}\delta^{(3)}(\mathbf{p}_{1}-\mathbf{p}_{2})\,
\end{eqnarray}
the spin reduced density matrix becomes
\begin{eqnarray}
\rho_{\Lambda}&=&\displaystyle\sum_{\sigma_1,\sigma_2,\sigma'_1,\sigma'_2}\int d^3\mathbf{p}\,\bigl|\psi(\mathbf{p})\bigr|^{2}\;c_{\sigma_1}c^{\ast}_{\sigma_2}\;D_{\sigma'_1\sigma_1}(W(\Lambda,\mathbf{p}))\nonumber\\
&\times &\bigl|\sigma'_{1}\bigr\rangle\bigl\langle\sigma'_{2}\bigr|\;D_{\sigma'_2\sigma_2}^{\dagger}(W(\Lambda,\mathbf{p}))\,.\label{boostedrho}
\end{eqnarray}
\section{Bipartite quantum state under Lorentz boost}\label{sec:boosted_state}\label{S3}
\noindent The expression in eq.~\eqref{boostedpsi} demonstrates that, for a single-particle state, a Lorentz boost intertwines spin and momentum via a momentum-dependent rotation which is often referred to as spin–momentum entanglement \cite{weinberg1995quantum}. In the following, we extend this analysis to a bipartite system. We ask how such entanglement influences the transformed state when an initially entangled two spin-$1/2$ particle state is subjected to boosts in two distinct settings. In the first scenario, only one of the particles experiences the Lorentz transformation; in the second, both particles are boosted simultaneously. For the entire calculation, we consider that the particles are moving along the $\hat{x}$-direction in their respective frame, and the Lorentz boost is applied along the $\hat{z}$-direction with respect to an inertial observer $\mathcal{O}$.
\subsection{When only one particle is under the Lorentz boost}
\noindent In this subsection, we consider that both the spin-$1/2$ particles are initially moving along the $\hat{x}$-direction with respect to an inertial observer $\mathcal{O}$.
Therefore, with respect to that inertial observer $\mathcal{O}$, the initial quantum state of the two spin-$1/2$ particles can be written as 
\begin{equation}\label{psi_inertial}
\begin{split}
|\Psi\rangle=&\int dp_{1}\,dp_{2}\;\psi(p_{1})\,\psi(p_{2})\;\bigl|p_{1},p_{2}\bigr\rangle\\&\;\otimes\;\Bigl(\sin\theta\; \bigl|01\bigr\rangle+\cos\theta\;\bigl|10\bigr\rangle\Bigr)~.
\end{split}
\end{equation}
Now if we placed one particle among this entangled pair in a moving frame $\mathcal{O'}$, where $\mathcal{O'}$ is connected with the inertial frame $\mathcal{O}$ through a Lorentz transformation $\Lambda$, then with respect to the moving frame $\mathcal{O'}$, the above quantum state can be written as 
\begin{align}
|\Psi_{\Lambda}\rangle
&=\bigl(U_1(\Lambda)\otimes\mathbb{I}_2\bigr)\,|\Psi\rangle \\[4pt]
&=\int dp_{1}\,dp_{2}\;\psi(p_{1})\,\psi(p_{2})
\;\sqrt{\frac{(\Lambda p_{1})^{0}}{p_{1}^{0}}}
\;\bigl|\Lambda p_{1},\,p_{2}\bigr\rangle
\nonumber\\
&\;\otimes\;\Bigl[
\sin\theta\;\bigl(D(W(\Lambda,p_{1}))\otimes\mathbb{I}\bigr)\,\bigl|0,1\bigr\rangle \nonumber\\
&\;+\;
\cos\theta\;\bigl(D(W(\Lambda,p_{1}))\otimes\mathbb{I}\bigr)\,\bigl|1,0\bigr\rangle
\Bigr]~.
\end{align}
Now we consider the four-momentum of the particle $p^{\mu}=(m \cosh \chi,\, m \sinh \chi\,\hat{x})$ and the velocity of the moving frame $\mathcal{O'}$, $\vec{v}=\tanh\alpha\, \hat{z}$. Therefore, setting $\hat{e}=\hat{z}$, and $\hat{f}=\hat{x}$ in the eq.(s)(\ref{little_group}, \ref{components}), we get
\begin{equation}
D(W(\Lambda,p_1))=\cos \frac{\phi}{2}\mathds{1}+i\sin\frac{\phi}{2}(\Sigma_2)\,\label{D(1+1)}
\end{equation}
with
\begin{eqnarray}
\cos\frac{\phi}{2}&=&\frac{\cosh\frac{\alpha}{2}\cosh\frac{\chi}{2}}{\sqrt{\frac{1}{2}+\frac{1}{2}\cosh\alpha\cosh\chi}}\nonumber\\
\sin\frac{\phi}{2}&=&\frac{\sinh\frac{\alpha}{2}\sinh\frac{\chi}{2}}{\sqrt{\frac{1}{2}+\frac{1}{2}\cosh\alpha\cosh\chi}}\label{comp(1+1)}~.
\end{eqnarray}
Using the following results
\begin{align}
    D(W(\Lambda,p_{1}))\bigl|0\bigr\rangle&=\cos\frac{\phi}{2}\bigl|0\bigr\rangle-\sin\frac{\phi}{2}\bigl|1\bigr\rangle\\
    D(W(\Lambda,p_{1}))\bigl|1\bigr\rangle&=\sin\frac{\phi}{2}\bigl|0\bigr\rangle+\cos\frac{\phi}{2}\bigl|1\bigr\rangle
\end{align}
we get
\begin{align}
|\Psi_{\Lambda}\rangle
&=\int dp_{1}\,dp_{2}\;\psi(p_{1})\,\psi(p_{2})
\;\sqrt{\frac{(\Lambda p_{1})^{0}}{p_{1}^{0}}}
\;\bigl|\Lambda p_{1},\,p_{2}\bigr\rangle
\nonumber\\
&\;\otimes\;\left[\sin\theta\;\cos\frac{\phi}{2}\,\bigl|0,1\bigr\rangle-\sin\theta\;\sin\frac{\phi}{2}\,\bigl|1,1\bigr\rangle\right.\nonumber\\
&\;+\;
\left.\cos\theta\;\sin\frac{\phi}{2}\,\bigl|0,0\bigr\rangle+\cos\theta\;\cos\frac{\phi}{2}\,\bigl|1,0\bigr\rangle\right]~.
\end{align}
The density matrix corresponding to the above state is given by
\begin{align}\label{rho_boosted}
&\varrho_{\Lambda}=\int dp_1\,dp_2\,dp'_1\,dp'_2\;
\psi(p_1)\,\psi(p_2)\,\psi^*(p'_1)\,\psi^*(p'_2)\nonumber\\
&\quad\times\sqrt{\frac{(\Lambda p_1)^0\,(\Lambda p'_1)^0}{p_1^0\,p_1'^0}}
\bigl|\Lambda p_1,p_2\bigr\rangle\bigl\langle\Lambda p'_1,p'_2\bigr|\;\otimes\;\nonumber\\
&\quad\Bigl[ \;\mathcal{C}\,\mathcal{C}^*\,|00\rangle\langle00|
\;+\;\mathcal{C}\,\mathcal{A}^*\,|00\rangle\langle01|
\;+\;\mathcal{C}\,\mathcal{D}^*\,|00\rangle\langle10|\nonumber\\
&\quad+\;\mathcal{C}\,\mathcal{B}^*\,|00\rangle\langle11|
+\,\mathcal{A}\,\mathcal{C}^*\,|01\rangle\langle00|
\;+\;\mathcal{A}\,\mathcal{A}^*\,|01\rangle\langle01|\nonumber\\
&\quad+\;\mathcal{A}\,\mathcal{D}^*\,|01\rangle\langle10|
\;+\;\mathcal{A}\,\mathcal{B}^*\,|01\rangle\langle11|\;
+\;\mathcal{D}\,\mathcal{C}^*\,|10\rangle\langle00|\nonumber\\
&\quad+\;\mathcal{D}\,\mathcal{A}^*\,|10\rangle\langle01|
\;+\;\mathcal{D}\,\mathcal{D}^*\,|10\rangle\langle10|
\;+\;\mathcal{D}\,\mathcal{B}^*\,|10\rangle\langle11|\nonumber\\
&\quad+\,\mathcal{B}\,\mathcal{C}^*\,|11\rangle\langle00|
\;+\;\mathcal{B}\,\mathcal{A}^*\,|11\rangle\langle01|
\;+\;\mathcal{B}\,\mathcal{D}^*\,|11\rangle\langle10|\nonumber\\
&\quad+\;\mathcal{B}\,\mathcal{B}^*\,|11\rangle\langle11|\Bigr]~.
\end{align}
where
\begin{align}
\mathcal{A}&=\sin\theta\;\cos\frac{\phi}{2},\quad
\mathcal{B}=-\sin\theta\;\sin\frac{\phi}{2}\nonumber\\
\mathcal{C}&=\cos\theta\;\sin\frac{\phi}{2},\quad
\mathcal{D}=\cos\theta\;\cos\frac{\phi}{2}~.
\end{align}
Taking the trace over the momentum degrees of freedom, the reduced density matrix corresponding to $\varrho_{\Lambda}$ becomes
\begin{align}\label{rho_reduced}
&\rho^{R}_{\Lambda}=\int dp_1\,dp_2\;
\bigl|\psi(p_1)\bigr|^{2}\,\bigl|\psi(p_2)\bigr|^{2}\;\otimes\;\Bigl[ \;\mathcal{C}\,\mathcal{C}^*\,|00\rangle\langle00|\nonumber\\
&\quad
\;+\;\mathcal{C}\,\mathcal{A}^*\,|00\rangle\langle01|
\;+\;\mathcal{C}\,\mathcal{D}^*\,|00\rangle\langle10|+\;\mathcal{C}\,\mathcal{B}^*\,|00\rangle\langle11|\nonumber\\
&\quad
\;+\,\mathcal{A}\,\mathcal{C}^*\,|01\rangle\langle00|
\;+\;\mathcal{A}\,\mathcal{A}^*\,|01\rangle\langle01|+\;\mathcal{A}\,\mathcal{D}^*\,|01\rangle\langle10|\nonumber\\
&\quad
\;+\;\mathcal{A}\,\mathcal{B}^*\,|01\rangle\langle11|\;
+\;\mathcal{D}\,\mathcal{C}^*\,|10\rangle\langle00|+\;\mathcal{D}\,\mathcal{A}^*\,|10\rangle\langle01|\nonumber\\
&\quad
\;+\;\mathcal{D}\,\mathcal{D}^*\,|10\rangle\langle10|
\;+\;\mathcal{D}\,\mathcal{B}^*\,|10\rangle\langle11|+\,\mathcal{B}\,\mathcal{C}^*\,|11\rangle\langle00|\nonumber\\
&\quad
\;+\;\mathcal{B}\,\mathcal{A}^*\,|11\rangle\langle01|
\;+\;\mathcal{B}\,\mathcal{D}^*\,|11\rangle\langle10|+\;\mathcal{B}\,\mathcal{B}^*\,|11\rangle\langle11\Bigr]~.
\end{align}
As $\sin\frac{\phi}{2}$, and $\cos\frac{\phi}{2}$ depends only on $p_1$, therefore we take
\begin{eqnarray}
    \int_{-\infty}^{\infty} dp_2\;
\bigl|\psi(p_2)\bigr|^{2}=1~.
\end{eqnarray}
Now defining the following integrals

\begin{align}
\mathcal{I}_1&=\int_{-\infty}^{\infty} dp_1\;
\bigl|\psi(p_1)\bigr|^{2}\cos^{2}\frac{\phi}{2}\label{def_int}
\end{align}

\begin{align}
\mathcal{I}_2&=\int_{-\infty}^{\infty} dp_1\;
\bigl|\psi(p_1)\bigr|^{2}\sin\frac{\phi}{2}\cos\frac{\phi}{2}\label{def_int2}\\
\mathcal{I}_3&=\int_{-\infty}^{\infty} dp_1\;
\bigl|\psi(p_1)\bigr|^{2}\sin^{2}\frac{\phi}{2}\label{def_int3}
\end{align}
we can rewrite the reduced density matrix in the form
\begin{widetext}
    \begin{eqnarray}\label{rho_final}
 \rho^{R}_{\Lambda}=
        \begin{pmatrix}
\cos^{2}\theta\,\mathcal{I}_3 & \sin\theta\cos\theta\,\mathcal{I}_2 & \cos^{2}\theta\,\mathcal{I}_2 & -\sin\theta\cos\theta\,\mathcal{I}_3\\
\sin\theta\cos\theta\,\mathcal{I}_2 & \sin^{2}\theta\,\mathcal{I}_1 & \sin\theta\cos\theta\,\mathcal{I}_1 & -\sin^{2}\theta\,\mathcal{I}_2\\
\cos^{2}\theta\,\mathcal{I}_2 & \sin\theta\cos\theta\,\mathcal{I}_1 & \cos^{2}\theta\,\mathcal{I}_1 & -\sin\theta\cos\theta\,\mathcal{I}_2\\
-\sin\theta\cos\theta\,\mathcal{I}_3 & -\sin^{2}\theta\,\mathcal{I}_2 &-\sin\theta\cos\theta\,\mathcal{I}_2 & \sin^{2}\theta\,\mathcal{I}_3
        \end{pmatrix}
        ~.
    \end{eqnarray}
\end{widetext}
To compute the components of the reduced density matrix $\rho^{R}_{\Lambda}$, we take the $\psi(p_1)$ in the following form
\begin{eqnarray}\label{form_psi}
\psi(p_1)=\frac{1}{\sqrt{\sigma^{2n+1}\Gamma(n+\frac{1}{2})}}p_1^{n}\,\mathrm{e}^{-\frac{1}{2}\left(\frac{p_1}{\sigma}\right)^2}\label{psi_1}
\end{eqnarray}
where $n\in\mathbb{R}$.
In order to represent $\sin\frac{\phi}{2}$ and $\cos\frac{\phi}{2}$ in terms of $p_1$, we take
\begin{align}
    \sinh\chi&=\frac{p_1}{m},\quad\cosh\chi=\sqrt{1+\left(\frac{p_1}{m}\right)^2}\nonumber\\
    \sinh\alpha&=a,\quad\cosh\alpha=b~. 
\end{align}
In terms of $a,\,b$, and $p_1$ we get
\begin{align}\label{form_comp}
    \cos^{2}\frac{\phi}{2}&=\frac{(1+b)\bigl(1+\sqrt{1+\left(\frac{p_1}{m}\right)^2}\bigr)}{2\,(1+b\sqrt{1+\left(\frac{p_1}{m}\right)^2})}\\
    \sin^{2}\frac{\phi}{2}&=\frac{(1-b)(1-\sqrt{1+\left(\frac{p_1}{m}\right)^2}}{2\,(1+b\sqrt{1+\left(\frac{p_1}{m}\right)^2})}\\
    \sin\frac{\phi}{2}\;\cos\frac{\phi}{2}&=\frac{a\frac{p_1}{m}}{2\,(1+b\sqrt{1+\left(\frac{p_1}{m}\right)^2})}~.
\end{align}
Using eq.(s)(\ref{form_psi},\ref{form_comp}) into eq.~\eqref{def_int}, we get
\begin{align}\label{int_1}
    \mathcal{I}_1&=\int_{-\infty}^{\infty} dp_1\;\frac{1}{\sigma^{2n+1}\Gamma(n+\frac{1}{2})}\;p_1^{2n}\;\mathrm{e}^{-\left(\frac{p_1}{\sigma}\right)^2}\nonumber\\
    &\quad\times\frac{(1+b)\bigl(1+\sqrt{1+\left(\frac{p_1}{m}\right)^2}\bigr)}{2\,(1+b\sqrt{1+\left(\frac{p_1}{m}\right)^2})}~.
\end{align}
Defining a new variable $\kappa=\frac{p_1}{\sigma}$ and recasting eq.~\eqref{int_1}, we get
\begin{align}\label{int_1a}
    \mathcal{I}_1&=\frac{(b+1)}{2\;\Gamma(n+\frac{1}{2})}\int_{-\infty}^{\infty} d\kappa\;\kappa^{2n}\;\mathrm{e}^{-\kappa^2}\;\frac{\bigl(1+\sqrt{1+\kappa^2\left(\frac{\sigma}{m}\right)^2}\bigr)}{(1+b\sqrt{1+\kappa^2\left(\frac{\sigma}{m}\right)^2})}~.
\end{align}
To perform the above integration, we consider the width of the wave function to be very small compared to the mass of the particle, as the exact analytical solution is hard to find. Therefore, in the limit $\left(\frac{\sigma}{m}\right)\ll1$, up to $\mathcal{O}\left[\left(\frac{\sigma}{m}\right)^2\right]$, we get
\begin{eqnarray}
    \mathcal{I}_1=\left\{\frac{1+(-1)^{2n}}{2}\right\}\left[1-\left(\frac{2n+1}{8}\right)\left(\frac{\cosh\alpha-1}{\cosh\alpha+1}\right)\left(\frac{\sigma}{m}\right)^2\right]~.\nonumber\\
\end{eqnarray}
Similarly, following the same approach, the other integrals can be evaluated and takes the form up to $\mathcal{O}\left[\left(\frac{\sigma}{m}\right)^2\right]$
\begin{eqnarray}
    \mathcal{I}_2=\left\{\frac{1-(-1)^{2n}}{2}\right\}\left[\frac{\Gamma(n+1)}{\Gamma(n+\frac{1}{2})}\frac{\sinh\alpha}{2\;(\cosh\alpha+1)}\left(\frac{\sigma}{m}\right)\right]~.\nonumber\\
\end{eqnarray}
\begin{eqnarray}
    \mathcal{I}_3=\left\{\frac{1+(-1)^{2n}}{2}\right\}\left[\left(\frac{2n+1}{8}\right)\left(\frac{\cosh\alpha-1}{\cosh\alpha+1}\right)\left(\frac{\sigma}{m}\right)^2\right]~.\nonumber\\
\end{eqnarray}
Taking the integer values of $n$, we get
\begin{eqnarray}
    \left\{\frac{1+(-1)^{2n}}{2}\right\}=1,\;\quad\text{and}\;\quad\left\{\frac{1-(-1)^{2n}}{2}\right\}=0~.\,\,\,\quad
\end{eqnarray}
Therefore, for the integer values of $n$, $\mathcal{I}_1$, $\mathcal{I}_2$, and $\mathcal{I}_3$ takes the form
\begin{eqnarray}\label{int_final}
    \mathcal{I}_1&=&(1-\mathcal{F}),\quad
    \mathcal{I}_2=0\nonumber\\
    \mathcal{I}_3&=&\mathcal{F}
\end{eqnarray}
where
\begin{eqnarray}
    \mathcal{F}=\left(\frac{2n+1}{8}\right)\left(\frac{\cosh\alpha-1}{\cosh\alpha+1}\right)\left(\frac{\sigma}{m}\right)^2~.
\end{eqnarray}
Putting the values of $\mathcal{I}_1$, $\mathcal{I}_2$, and $\mathcal{I}_3$ given in eq.~\eqref{int_final} in eq.~\eqref{rho_final}, the reduced density matrix turns out to be
\begin{widetext}
    \begin{eqnarray}\label{rho_final1}
 \rho^{R}_{\Lambda}=
        \begin{pmatrix}
\cos^{2}\theta\,\mathcal{F} & 0 & 0 & -\sin\theta\cos\theta\,\mathcal{F}\\
0 & \sin^{2}\theta\,(1-\mathcal{F}) & \sin\theta\cos\theta\,(1-\mathcal{F}) & 0\\
0 & \sin\theta\cos\theta\,(1-\mathcal{F}) & \cos^{2}\theta\,(1-\mathcal{F}) & 0\\
-\sin\theta\cos\theta\,\mathcal{F} & 0 & 0 & \sin^{2}\theta\,\mathcal{F}
        \end{pmatrix}
        ~.
    \end{eqnarray}
\end{widetext}
\subsection{When both the particles are under the Lorentz boost}
\noindent In this subsection, we study the scenario where each spin-$1/2$ particle resides in its own moving frame, denoted $\mathcal{O}'$ and $\mathcal{O}''$. Frame $\mathcal{O}'$ travels at velocity $v_1$ and is related to the inertial frame $\mathcal{O}$ by the Lorentz transformation $\Lambda_1$, while $\mathcal{O}''$ moves at velocity $v_2$ and connects to $\mathcal{O}$ via $\Lambda_2$.
The initial quantum state of the two spin-$1/2$ particles in this scenario will remain identical to eq.~\eqref{psi_inertial}.

In the moving frames, the two-particle quantum state can be written as 
\begin{align}
&\quad\bigl|\Psi_{\Lambda_1\Lambda_2}\bigr\rangle
=\bigl(U_1(\Lambda_1)\otimes U_2(\Lambda_2)\bigr)\,\bigl|\Psi\bigr\rangle \\[4pt]
&=\int dp_{1}\,dp_{2}\;\psi(p_{1})\,\psi(p_{2})
\;\sqrt{\frac{(\Lambda p_{1})^{0}(\Lambda p_{2})^{0}}{p_{1}^{0}p_{2}^{0}}}
\;\bigl|\Lambda_{1}p_{1},\,\Lambda_{2}p_{2}\bigr\rangle
\nonumber\\
&\;\otimes\;\Bigl[
\sin\theta\;\bigl(D(W(\Lambda_{1},p_{1}))\otimes D(W(\Lambda_{2},p_{2}))\bigr)\,\bigl|0,1\bigr\rangle \nonumber\\
&\;+\;
\cos\theta\;\bigl(D(W(\Lambda_{1},p_{1}))\otimes D(W(\Lambda_{2},p_{2}))\bigr)\,\bigl|1,0\bigr\rangle
\Bigr]~.
\end{align}
Using the following results
\begin{align}
    D(W(\Lambda_1,p_{1}))\bigl|0\bigr\rangle&=\cos\frac{\phi_1}{2}\bigl|0\bigr\rangle-\sin\frac{\phi_1}{2}\bigl|1\bigr\rangle\\
    D(W(\Lambda_1,p_{1}))\bigl|1\bigr\rangle&=\sin\frac{\phi_1}{2}\bigl|0\bigr\rangle+\cos\frac{\phi_1}{2}\bigl|1\bigr\rangle\\
    D(W(\Lambda_2,p_{2}))\bigl|0\bigr\rangle&=\cos\frac{\phi_2}{2}\bigl|0\bigr\rangle-\sin\frac{\phi_2}{2}\bigl|1\bigr\rangle\\
    D(W(\Lambda_2,p_{2}))\bigl|1\bigr\rangle&=\sin\frac{\phi_2}{2}\bigl|0\bigr\rangle+\cos\frac{\phi_2}{2}\bigl|1\bigr\rangle
\end{align}
the two-particle quantum state becomes
\begin{align}\label{psi_boosted_two}
&\quad\bigl|\Psi_{\Lambda_1\Lambda_2}\bigr\rangle
\\[4pt]
&=\int dp_{1}\,dp_{2}\;\psi(p_{1})\,\psi(p_{2})
\;\sqrt{\frac{(\Lambda p_{1})^{0}(\Lambda p_{2})^{0}}{p_{1}^{0}p_{2}^{0}}}
\;\bigl|\Lambda_{1}p_{1},\,\Lambda_{2}p_{2}\bigr\rangle
\nonumber\\
&\;\otimes\;\Bigl[
\mathcal{P}\,\bigl|0,0\bigr\rangle\;+\;\mathcal{Q}\,\bigl|0,1\bigr\rangle\;+\;\mathcal{R}\,\bigl|1,0\bigr\rangle\;+\;\mathcal{S}\,\bigl|1,1\bigr\rangle\Bigr]~.
\end{align}
where
\begin{align}\label{P,Q,R,S}
    \mathcal{P}&=\left(\sin\theta\cos\frac{\phi_1}{2}\sin\frac{\phi_2}{2}+\cos\theta\sin\frac{\phi_1}{2}\cos\frac{\phi_2}{2}\right)\nonumber\\
    \mathcal{Q}&=\left(\sin\theta\cos\frac{\phi_1}{2}\cos\frac{\phi_2}{2}-\cos\theta\sin\frac{\phi_1}{2}\sin\frac{\phi_2}{2}\right)\nonumber\\
    \mathcal{R}&=-\left(\sin\theta\sin\frac{\phi_1}{2}\sin\frac{\phi_2}{2}-\cos\theta\cos\frac{\phi_1}{2}\cos\frac{\phi_2}{2}\right)\nonumber\\
    \mathcal{S}&=-\left(\sin\theta\sin\frac{\phi_1}{2}\cos\frac{\phi_2}{2}+\cos\theta\cos\frac{\phi_1}{2}\sin\frac{\phi_2}{2}\right)~.
\end{align}
The density matrix corresponding to the state given in eq.~\eqref{psi_boosted_two} takes the form
\begin{widetext}
\begin{align}\label{rho_boosted_two}
&\varrho_{\Lambda_1\Lambda_2}=\int dp_1\,dp_2\,dp'_1\,dp'_2\;
\psi(p_1)\,\psi(p_2)\,\psi^*(p'_1)\,\psi^*(p'_2)\sqrt{\frac{(\Lambda_1 p_1)^0\,(\Lambda_2 p_2)^0\,(\Lambda_1 p'_1)^0\,(\Lambda_2 p'_2)^0}{p_1^0\,p_2^0\,p_1'^0\,p_2'^0}}
\bigl|\Lambda_1 p_1,\Lambda_2 p_2\bigr\rangle\bigl\langle\Lambda_1 p'_1,\Lambda_2 p'_2\bigr|\nonumber\\[4pt]
&\quad\otimes\;\Bigl[
\mathcal{P}\,\bigl|0,0\bigr\rangle\;+\;\mathcal{Q}\,\bigl|0,1\bigr\rangle\;+\;\mathcal{R}\,\bigl|1,0\bigr\rangle\;+\;\mathcal{S}\,\bigl|1,1\bigr\rangle\Bigr]\Bigl[\mathcal{P}\,\bigl\langle0,0\bigr|\;+\;\mathcal{Q}\,\bigl\langle0,1\bigr|\;+\;\mathcal{Q}\,\bigl\langle1,0\bigr|\;+\;\mathcal{Q}\,\bigl\langle1,1\bigr|\Bigr]~.
\end{align}
\end{widetext}
Tracing over the momentum degrees of freedom of both the particles, the reduced density matrix corresponding to the density matrix of the two-particle system $\varrho_{\Lambda_1\Lambda_2}$ becomes
\begin{align}\label{rho_reduced_two}
\rho^{R}_{\Lambda_1\Lambda_2}&=\int dp_1\,dp_2\;
\bigl|\psi(p_1)\bigr|^{2}\,\bigl|\psi(p_2)\bigr|^{2}\nonumber\\
&\quad\otimes\;\Bigl[
\mathcal{P}\,\bigl|0,0\bigr\rangle\;+\;\mathcal{Q}\,\bigl|0,1\bigr\rangle\;+\;\mathcal{R}\,\bigl|1,0\bigr\rangle\;+\;\mathcal{S}\,\bigl|1,1\bigr\rangle\Bigr]\nonumber\\
&\quad\times\;\Bigl[\mathcal{P}\,\bigl\langle0,0\bigr|\;+\;\mathcal{Q}\,\bigl\langle0,1\bigr|\;+\;\mathcal{Q}\,\bigl\langle1,0\bigr|\;+\;\mathcal{Q}\,\bigl\langle1,1\bigr|\Bigr]~.
\end{align}
We now define the following integrals
\begin{align}\label{def_int_f}
\mathcal{J}_i&=\int_{-\infty}^{\infty} dp_i\;
\bigl|\psi(p_i)\bigr|^{2}\cos^{2}\frac{\phi_i}{2}\\
\mathcal{K}_i&=\int_{-\infty}^{\infty} dp_i\;
\bigl|\psi(p_i)\bigr|^{2}\sin\frac{\phi_i}{2}\cos\frac{\phi_i}{2}\\
\mathcal{L}_i&=\int_{-\infty}^{\infty} dp_i\;
\bigl|\psi(p_i)\bigr|^{2}\sin^{2}\frac{\phi_i}{2}
\end{align}
where $i=1,\,2$~. For computing the components of the reduced density matrix $\rho^{R}_{\Lambda_1\Lambda_2}$, we take $\psi(p_i)$ in the following form
\begin{eqnarray}\label{form_psi_two}
\psi(p_i)=\frac{1}{\sqrt{\sigma^{2n+1}\Gamma(n+\frac{1}{2})}}p_i^{n}\,\mathrm{e}^{-\frac{1}{2}\left(\frac{p_i}{\sigma}\right)^2}\label{psi_i}
\end{eqnarray}
where $n\in\mathbb{R}$.
In order to represent $\sin\frac{\phi_i}{2}$ and $\cos\frac{\phi_i}{2}$ in terms of $p_i$, we take
\begin{align}
    \sinh\chi_i&=\frac{p_i}{m},\quad\cosh\chi_i=\sqrt{1+\left(\frac{p_i}{m}\right)^2}\nonumber\\
    \sinh\alpha_i&=a_i,\quad\cosh\alpha_i=b_i~. 
\end{align}
In terms of $a_i,\,b_i$, and $p_i$, we get
\begin{align}
    \cos^{2}\frac{\phi_i}{2}&=\frac{(1+b_i)\bigl(1+\sqrt{1+\left(\frac{p_i}{m}\right)^2}\bigr)}{2\,(1+b_i\sqrt{1+\left(\frac{p_i}{m}\right)^2})}\label{form_comp_two1}\\
    \sin^{2}\frac{\phi_i}{2}&=\frac{(1-b_i)(1-\sqrt{1+\left(\frac{p_i}{m}\right)^2}}{2\,(1+b_i\sqrt{1+\left(\frac{p_i}{m}\right)^2})}\label{form_comp_two2}\\
    \sin\frac{\phi_i}{2}\;\cos\frac{\phi_i}{2}&=\frac{a_i\frac{p_i}{m}}{2\,(1+b_i\sqrt{1+\left(\frac{p_i}{m}\right)^2})}\label{form_comp_two3}
\end{align}
with $i=1,2$.
Using the form of $\psi(p_1)$ and $\psi(p_2)$ given in eq.~\eqref{psi_i} by taking integer values of $n$ and the form given in eq(s).~(\ref{form_comp_two1}, \ref{form_comp_two2}, \ref{form_comp_two3}), we evaluate the integrals $\mathcal{J}_i$, $\mathcal{K}_i$, and $\mathcal{L}_i$, and the results are given by
\begin{eqnarray}\label{J,K,L}
    \mathcal{J}_1&=&(1-\mathcal{F}_1),\quad\mathcal{J}_2=(1-\mathcal{F}_2),\quad
    \mathcal{K}_1=\mathcal{K}_2=0\nonumber\\
\mathcal{L}_1&=&\mathcal{F}_1,\quad\mathcal{L}_2=\mathcal{F}_2
\end{eqnarray}
with
\begin{eqnarray}\label{F_i}
    \mathcal{F}_i=\left(\frac{2n+1}{8}\right)\left(\frac{\cosh\alpha_i-1}{\cosh\alpha_i+1}\right)\left(\frac{\sigma}{m}\right)^2~.
\end{eqnarray}
Putting the values given in eq(s).~(\ref{J,K,L}, \ref{F_i}) into eq.~\eqref{rho_reduced_two}, we can recast the reduced density matrix $\rho^{R}_{\Lambda_1\Lambda_2}$ as
\begin{widetext}
    \begin{eqnarray}\label{rho_final1_two}
 \rho^{R}_{\Lambda_1\Lambda_2}=
        \begin{pmatrix}
\sin^{2}\theta\,\mathcal{F}_1+\cos^{2}\theta\,\mathcal{F}_2 & 0 & 0 & -\sin\theta\cos\theta\,(\mathcal{F}_1+\mathcal{F}_2)\\
0 & \sin^{2}\theta\,(1-\mathcal{F}_1-\mathcal{F}_2) & \sin\theta\cos\theta\,(1-\mathcal{F}_1-\mathcal{F}_2) & 0\\
0 & \sin\theta\cos\theta\,(1-\mathcal{F}_1-\mathcal{F}_2) & \cos^{2}\theta\,(1-\mathcal{F}_1-\mathcal{F}_2) & 0\\
-\sin\theta\cos\theta\,(\mathcal{F}_1+\mathcal{F}_2) & 0 & 0 & \sin^{2}\theta\,\mathcal{F}_2+\cos^{2}\theta\,\mathcal{F}_1
        \end{pmatrix}
        ~.\quad\quad
    \end{eqnarray}
\end{widetext}
\section{Effect of the Lorentz boost on quantum coherence measures}\label{S4}
In this section, we will calculate different measures of quantum coherence. As quantum coherence is one of the important quantum resources, therefore, it is intriguing to see the effect of relativistic motion on quantum coherence in the entanglement set up.
\subsection{$l_1$ norm measure of coherence}
\noindent Here, we compute the $l_1$-norm coherence measure, defined as \cite{l1_Norm}
\begin{eqnarray}
    C_{l_1}\bigl(\rho^{R}_{\Lambda} \bigr)
&= \displaystyle\sum_{\substack{i,j=1 \\ i\neq j}}^{4} \bigl|\rho_{ij}\bigr|~.
\end{eqnarray}
From eq.~\eqref{rho_final1}, we observe that the only nonzero off–diagonal entries of $\rho^{R}_{\Lambda}$ are
\begin{eqnarray}
    \rho_{14}&=&\rho_{41}=-\sin\theta\cos\theta\,\mathcal{F}\nonumber\\
    \rho_{23}&=&\rho_{32}=\sin\theta\cos\theta\,(1-\mathcal{F})~.
\end{eqnarray}
Therefore,
\begin{align*}
C_{l_1}\bigl(\rho^{R}_{\Lambda}\bigr)
&=\sum_{i\neq j}\bigl|\rho_{ij}\bigr|\\&
=|\rho_{14}|+|\rho_{41}|+|\rho_{23}|+|\rho_{32}|\\
&=2\bigl|\sin\theta\cos\theta\,\mathcal{F}\bigr|
+2\bigl|\sin\theta\cos\theta\,(1-\mathcal{F})\bigr|\\
&=2\bigl|\sin\theta\cos\theta\bigr|\,
\bigl(\mathcal{F}+(1-\mathcal{F})\bigr)
=2\bigl|\sin\theta\cos\theta\bigr|\,.
\end{align*}
In particular, if \(0\le\theta\le\pi/2\), this further reduces to
\begin{eqnarray}\label{l_1 measure}
C_{l_1}\bigl(\rho^{R}_{\Lambda}\bigr)
=2\sin\theta\cos\theta
=\sin(2\theta)\,.
\end{eqnarray}
Using eq.~\eqref{rho_final1_two}, and following a similar approach, we observe that the only nonzero off–diagonal entries of $\rho^{R}_{\Lambda_1\Lambda_2}$ are
\begin{eqnarray}
    \rho_{14}&=&\rho_{41}=-\sin\theta\cos\theta\,\bigl(\mathcal{F}_1+\mathcal{F}_2\bigr)\nonumber\\
    \rho_{23}&=&\rho_{32}=\sin\theta\cos\theta\,\bigl[1-\bigl(\mathcal{F}_1+\mathcal{F}_2\bigr)\bigr]~.
\end{eqnarray}
Therefore,
\begin{align}\label{l_1 measure_two}
C_{l_1}\bigl(\rho^{R}_{\Lambda_1\Lambda_2}\bigr)
&=\sum_{i\neq j}\bigl|\rho_{ij}\bigr|
=|\rho_{14}|+|\rho_{41}|+|\rho_{23}|+|\rho_{32}|\nonumber\\
&=2\bigl|\sin\theta\cos\theta\bigr|\,
\bigl[\mathcal{F}_1+\mathcal{F}_2+\bigl(1-\mathcal{F}_1-\mathcal{F}_2\bigr)\bigr]\nonumber\\
&=2\bigl|\sin\theta\cos\theta\bigr|=\sin(2\theta)\,.
\end{align}
\subsection{Frobenius-norm measure of coherence}
\noindent The Frobenius-norm measure of coherence is given by \cite{YaoDongXiaoSun}
\begin{eqnarray}\label{def_Frob_measure}
C_{F}\bigl(\rho^{R}_{\Lambda}\bigr)\equiv\sqrt{\frac{d}{d-1}\displaystyle\sum_{i=1}^d \left(\lambda_{i}-\frac{1}{d}\right)^2}~.
\end{eqnarray}
Here, $d$ denotes the dimension of the Hilbert space, and the eigenvalues of the reduced density matrix $\rho^{R}_{\Lambda}$ are given by $\{\lambda_i\}$.
Calculating the eigenvalues of the reduced density matrix $\rho^{R}_{\Lambda}$ from eq.~\eqref{rho_final1}, we get
\begin{eqnarray}
    \{\lambda_i\}=\bigl\{\mathcal{F},\,(1-\mathcal{F}),\,0,\,0\bigr\}~.
\end{eqnarray}
Then using it in eq.~\eqref{def_Frob_measure}, and taking the approximation up to $\mathcal{O}\left[\left(\frac{\sigma}{m}\right)^2\right]$, we get
\begin{eqnarray}\label{Frob_measure}
    C_{F}\bigl(\rho^{R}_{\Lambda}\bigr)=\left[1-\left(\frac{2n+1}{6}\right)\left(\frac{\cosh\alpha-1}{\cosh\alpha+1}\right)\left(\frac{\sigma}{m}\right)^2\right]~.\quad\quad
\end{eqnarray}
Here, $\alpha$ is the rapidity parameter for the boosted observer, and $\cosh\alpha=\gamma\equiv 1/\sqrt{1-\frac{v^2}{c^2}}$, where $v$ is the velocity of the boosted observer with respect to the inertial one. Therefore, when $v\rightarrow 0$, $\cosh\alpha\rightarrow 1$. In this limit, eq.~\eqref{Frob_measure} becomes
\begin{eqnarray}
C_{F}(\rho^{R}_{\Lambda})=1\label{Frob_measure1}\,.
\end{eqnarray}
On the other hand, when $v\rightarrow c,~\text{ then }\cosh\alpha\rightarrow \infty$. In this limit, eq.~\eqref{Frob_measure} becomes
\begin{eqnarray}
C_{F}(\rho^{R}_{\Lambda})=\left[1-\frac{2n+1}{6}\left(\frac{\sigma}{m}\right)^2\right]\label{Frob_measure2}\,.
\end{eqnarray}
Since we must have $C_{F}(\rho^{R}_{\Lambda})\geq0$, hence we get an upper bound of the parameter $n$ to be
\begin{eqnarray}
n\leq \left[3\left(\frac{m}{\sigma}\right)^2-\frac{1}{2}\right]~.\label{bound_on_n}
\end{eqnarray}
As maximum coherence must be less than or equal to unity, therefore, using this condition, we fix the lower bound for $n$. Considering $\frac{\sigma}{m}<1$, we obtain
\begin{equation}\label{n_lb}
\begin{split}
&1-\frac{2n+1}{6}\left(\frac{\sigma}{m}\right)^2<1\\
\implies&n> -\frac{1}{2}~.
\end{split}
\end{equation}
Combining eq.(\ref{bound_on_n}) with eq.(\ref{n_lb}), we obtain a range for the parameter $n$ as
\begin{equation}\label{n_range}
-\frac{1}{2}< n\leq\left[3\left(\frac{m}{\sigma}\right)^2-\frac{1}{2}\right]~.
\end{equation}

\noindent Now calculating the eigenvalues of the reduced density matrix $\rho^{R}_{\Lambda_1\Lambda_2}$ from eq.~\eqref{rho_final1_two}, we get
\begin{align}
\xi_{1} &= \frac{\mathcal{F}_1 + \mathcal{F}_2}{2}
+ \frac{1}{2}\sqrt{\mathcal{F}_1^2 + \mathcal{F}_2^2 - 2\,\mathcal{F}_1\,\mathcal{F}_2\cos(4\theta)},\nonumber\\
\xi_{2} &= \frac{\mathcal{F}_1 + \mathcal{F}_2}{2}
- \frac{1}{2}\sqrt{\mathcal{F}_1^2 + \mathcal{F}_2^2 - 2\,\mathcal{F}_1\,\mathcal{F}_2\cos(4\theta)},\nonumber\\
\xi_{3} &= 1 - \bigl(\mathcal{F}_1 + \mathcal{F}_2\bigr),\nonumber\\
\xi_{4} &= 0.
\end{align}
Using the above eigenvalues in eq.~\eqref{def_Frob_measure}, and taking the approximation up to $\mathcal{O}\left[\left(\frac{\sigma}{m}\right)^2\right]$, we get
\begin{eqnarray}\label{Frob_measure_two}
    C_{F}\bigl(\rho^{R}_{\Lambda_1\Lambda_2}\bigr)&=&\left[1-\left(\frac{2n+1}{6}\right)\left(\frac{\cosh\alpha_1 -1}{\cosh\alpha_1 +1}\right)\left(\frac{\sigma}{m}\right)^2\right.\nonumber\\
    &\,&\left.\quad-\left(\frac{2n+1}{6}\right)\left(\frac{\cosh\alpha_2 -1}{\cosh\alpha_2 +1}\right)\left(\frac{\sigma}{m}\right)^2\right]~.\quad\quad
\end{eqnarray}
Here, $\cosh\alpha_1=1/\sqrt{1-\frac{v_1^2}{c^2}}$ and $\cosh\alpha_2=1/\sqrt{1-\frac{v_2^2}{c^2}}$, where $v_1$ and $v_2$ are the velocities of the boosted observers with respect to the inertial one. Therefore, when $v_1$ and $v_2$ both tend to $0$, then eq.~\eqref{Frob_measure_two} becomes
\begin{eqnarray}
C_{F}(\rho^{R}_{\Lambda_1\Lambda_2})=1\label{Frob_measure1_two}\,.
\end{eqnarray}
On the other hand, when $v_1$ and $v_2$ both tend to $c$, then eq.~\eqref{Frob_measure_two} becomes
\begin{eqnarray}
C_{F}(\rho^{R}_{\Lambda_1\Lambda_2})=\left[1-\frac{2n+1}{3}\left(\frac{\sigma}{m}\right)^2\right]\label{Frob_measure2_two}\,.
\end{eqnarray}
Again since we must have \(C_{F}(\rho^{R}_{\Lambda_1\Lambda_2})\ge0\), we immediately obtain an upper bound on \(n\), which reads
\begin{equation}
n \;\le\; \frac{3}{2}\Bigl(\frac{m}{\sigma}\Bigr)^{2} \;-\;\frac{1}{2}\,.
\end{equation}
Now, requiring that the maximum coherence does not exceed unity and assuming \(\frac{\sigma}{m}<1\),
\begin{equation}
1 \;-\;\frac{2n+1}{6}\Bigl(\frac{\sigma}{m}\Bigr)^{2}<1
\quad\Longrightarrow\quad
n>-\frac{1}{2}\,.
\end{equation}
Combining these two results, the allowed range for \(n\) is given by
\begin{equation}
-\tfrac{1}{2} \;<\; n \;\le\; \frac{3}{2}\Bigl(\frac{m}{\sigma}\Bigr)^{2} \;-\;\frac{1}{2}\,.
\end{equation}
\subsection{Reduction into a single-particle scenario}
\noindent In this subsection, we want to reduce the two-particle system into a single particle scenario so that we can compare the results of this work with the results of our previous work
\cite{MSG2024quantum}. For doing so, we are going to trace over the spin degrees of freedom of one particle.

At first we consider the reduced density matrix of the system where the Lorentz boost acts only on one particle among the entangled pair, given in eq.~\eqref{rho_final1}. Here, we trace out the spin degrees of freedom of the inertial particle.

After taking the trace over the spin degrees of freedom of the second particle, the reduced density matrix $\rho^{R}_{\Lambda}$ becomes

\begin{eqnarray}\label{rho_final_s}
 \rho^{R}_{\Lambda,\,\sigma_1}=
        \begin{pmatrix}
\sin^{2}\theta+\cos2\theta\,\mathcal{F} & 0 \\
 0 & \cos^{2}\theta-\cos2\theta\,\mathcal{F}
        \end{pmatrix}
        ~.\quad
    \end{eqnarray}
    
 \noindent From the above form of $\rho^{R}_{\Lambda,\,\sigma_1}$ it is observed that if we consider the maximally entangled state, then $\sin\theta=\cos\theta=1/\sqrt{2}$. For this case there will be no effect of the Lorentz boost on the system. For the non-maximal entangled state, the effecty of the Lorentz boost can be observed on the system.

 Comparing the above scenario with the scenario discussed in our previous work \cite{MSG2024quantum}, we observe that the structure of the density matrix $\rho^{R}_{\Lambda,\,\sigma_1}$ is completely different with the density matrix $\rho^{R}_{\Lambda}$ given in eq.~($40$) of \cite{MSG2024quantum}. 

 Now we consider the reduced density matrix of the system where the Lorentz boost is acted on both the particles given in eq.~\eqref{rho_final1_two}. In this scenario one can trace out the spin degrees of freedom of any one of the two particles to reduce the system into a single particle scenario.

 After tracing out the second particle, the reduced density matrix $\rho^{R}_{\Lambda_1\Lambda_2}$ becomes
 
\begin{eqnarray}\label{rho_final_s1}
 \rho^{R}_{\Lambda_1\Lambda_2,\,\sigma_1}=
        \begin{pmatrix}
\sin^{2}\theta+\cos2\theta\,\mathcal{F}_2 & 0 \\
 0 & \cos^{2}\theta-\cos2\theta\,\mathcal{F}_2
        \end{pmatrix}
        ~.\quad\quad
    \end{eqnarray}
    
\noindent While tracing out the first particle, the reduced density matrix $\rho^{R}_{\Lambda_1\Lambda_2}$ becomes

\begin{eqnarray}\label{rho_final_s2}
 \rho^{R}_{\Lambda_1\Lambda_2,\,\sigma_2}=
        \begin{pmatrix}
\cos^{2}\theta-\cos2\theta\,\mathcal{F}_1 & 0 \\
 0 & \sin^{2}\theta+\cos2\theta\,\mathcal{F}_1
        \end{pmatrix}
        ~.\quad\quad
    \end{eqnarray}
\section{Findings}\label{S5}
\noindent In this section, we shall show the effect of the Lorentz boost on the Frobenius-norm measure of coherence with respect to the change in Gaussian width for different values of the boost parameter.

To observe the effect of the Lorentz boost of the Frobenius-norm measure, we plot it with respect to the width of the Gaussian wave packet. Here, we consider the case of neutron which has a rest mass of $m=939.36$ MeV. It is observed that in the single‑particle Lorentz‑boost scenario, the neutron’s heavier rest‑mass energy leads to a more gradual decay of its coherence \cite{MSG2024quantum}.
\begin{figure}[h!]
\includegraphics[scale=0.9]{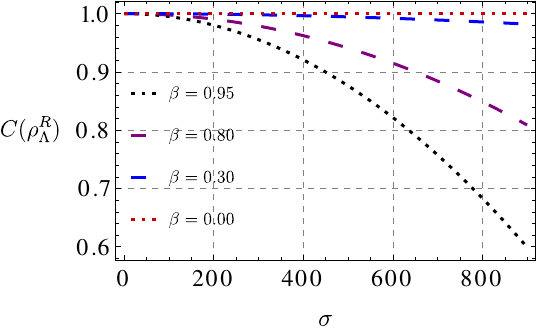}
\caption{The Frobenius norm measure of coherence, $C(\rho^{R}_\Lambda)$ plotted with the values of the boost parameter $\beta~=~0.95,~0.8,~0.3,~0.0$,  with $n=2$.}
\label{fig:F1}
\end{figure}
Fig. \ref{fig:F1} shows the behaviour of the Frobenius-norm measure for the case where only one particle of the entangle pair is affected by the Lorentz transformation. In Fig. \ref{fig:F1}, we observe that for an increase in the width of the Gaussian wave packet $\sigma$, the loss of
coherence is significant when the value of the boost parameter is very high, that is, $\beta= 0.95$. We have also seen that for lower values of the boost parameter $\beta$, the coherence loss is almost negligible. This kind of behaviour is consistent with the results given in Ref.~\cite{MSG2024quantum}.
\begin{figure}[ht!]
\includegraphics[scale=0.9]{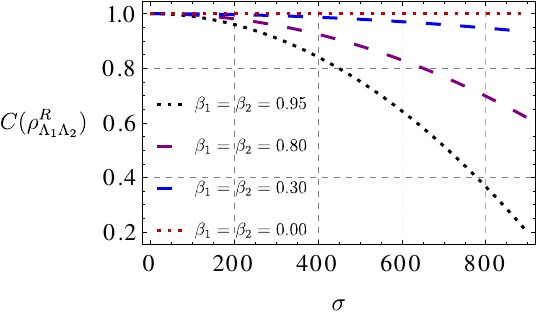}
\caption{The Frobenius norm measure of coherence, $C(\rho^{R}_{\Lambda_1\Lambda_2})$ plotted with the values of the boost parameter $\beta_1=\beta_2=0.95,~0.8,~0.3,~0.0$, with $n=2$.}
\label{fig:F2}
\end{figure}
In Figure~\ref{fig:F2}, we examine how coherence behaves when both members of our entangled pair are acted upon by Lorentz boosts. Fig.~\ref{fig:F2} shows that for small values of $\sigma$, the coherence remains relatively high across all boost settings. However, once $\sigma$ grows beyond a certain point, the curve corresponding to the maximal boost ($\beta_{1} = \beta_{2} = 0.95$) plunges much more steeply than the lower‑boost cases. Physically, this happens because each neutron’s large rest‑mass energy amplifies the relativistic spin–momentum coupling. When both particles are boosted, their individual coherence decays reinforce one another, producing a pronounced overall loss of coherence at moderate and large wave‑packet widths. Nevertheless, when the boost parameters are small in this configuration, the coherence degradation still remains minimal.
\section{Conclusions}\label{sec:Conclu}
\noindent To investigate the impact of Lorentz boost on quantum coherence measures in entangled systems, in this work, we consider an entangled bipartite system initially moving with respect to an inertial observer $\mathcal{O}$. In this inertial frame, we consider the initial state of the bipartite system in a generic entangled state. To see the effect of the Lorentz boost on the two particle quantum state of the system, we consider two scenarios in section \ref{sec:boosted_state}. In the first scenario, we consider that among the entangled pair, one particle is in the inertial frame while other particle is placed in a boosted reference frame. Whereas in the second scenario, we consider that both the particles are placed in different boosted frames, which are moving with, in general, different velocities with respect to the inertial observer $\mathcal{O}$. For both the cases, the boosted reference frames are moving to the perpendicular direction with the motion of the particles. For both the scenarios, we considered generalized form of the Gaussian wave packets with its peak at the origin. This generic Gaussian is multiplied by the $n^{th}$ power of the momentum $p$, and the wave function is in momentum space. 

After giving a Lorentz boost in a particular direction, we have calculated the two-particle entangled quantum states in the above two scenarios, which lead to the estimation of different measures of coherence with a dependence on the boost
parameter $\beta$, the Gaussian width $\sigma$, the mass of the particles, the generalization parameter $n$, and the entanglement parameter $\theta$. 

In this work, we mainly focused on two measures of coherence, namely, the $l_1$-norm measure of coherence and the Frobenius-norm measure of coherence with the assumption that $\left(\frac{\sigma}{m}\right)<1$. As we have considered that the Gaussian wave packets have narrow width, therefore both the measures of coherence are calculated up to $\mathcal{O}\left[\left(\frac{\sigma}{m}\right)^2\right]$. From the approximated results  of the $l_1$-norm measure of coherence, we have observed there is no effect of the boost parameter on the $l_1$-norm measure of coherence for the cases where the Lorentz boost acts on one or two particles of the entangled system. For a given entanglement parameter $\theta$, it remains constant with respect to the Gaussian width $\sigma$.

The differences between these two scenarios become noticeable when we see the behaviour of the Frobenius-norm measure of coherence. For the first scenario, the Frobenius-norm measure $C(\rho^{R}_{\Lambda})$ shows that when we increase the width of the Gaussian wave packet $\sigma$, the coherence loss is more prominent for the cases where the boost parameters takes high values such as $\beta>0.9$. In case of low boost value ($\beta=0.3$), the coherence loss remains insignificant. In the second scenario, the Frobenius-norm measure $C(\rho^{R}_{\Lambda_1\Lambda_2})$ shows that when we increase the width of the Gaussian wave packet $\sigma$, the coherence loss is much more stronger than that of previous case for the high values of boost parameters, $\beta_1=\beta_2=0.95$. Here also, the coherence loss remains negligible for the low boost values, $\beta_1=\beta_2=0.3$. Physically, this happens because each particle’s large rest-mass energy amplifies the relativistic spin–momentum coupling. When both particles are boosted, their individual coherence decays reinforce one another to produce a pronounced overall loss of coherence
at moderate and large wave-packet widths.

Finally, to compare the results of this work with our previous work \cite{MSG2024quantum}, we reduce the density matrices corresponding to the two particle boosted quantum states given in eq(s).~(\ref{boostedpsi}, \ref{psi_boosted_two}) into a single particle scenario by tracing the spin degrees of freedom of one particle. After the reduction, we conclude that to capture the Lorentz boost effect in a single particle scenario, the particles must be entangled non-maximally; otherwise, the relativistic effects would not be captured.
\bibliographystyle{hephys.bst}
\bibliography{Ref}

\begin{thebibliography}{10}
\newcommand{\enquote}[1]{``#1''}

\bibitem{EPR}
A.~Einstein, B.~Podolsky and N.~Rosen, \enquote{Can Quantum-Mechanical
  Description of Physical Reality Be Considered Complete?},
  \href{https://dx.doi.org/10.1103/PhysRev.47.777}{\emph{Phys. Rev.}
  \textbf{47} (1935) 777}.

\bibitem{Schrodinger_1935}
E.~Schrödinger, \enquote{Discussion of Probability Relations between Separated
  Systems},
  \href{https://dx.doi.org/10.1017/S0305004100013554}{\emph{Mathematical
  Proceedings of the Cambridge Philosophical Society} \textbf{31[4]} (1935)
  555–563}.

\bibitem{BellEPR}
J.~S. Bell, \enquote{On the Einstein Podolsky Rosen paradox},
  \href{https://dx.doi.org/10.1103/PhysicsPhysiqueFizika.1.195}{\emph{Physics
  Physique Fizika} \textbf{1} (1964) 195}.

\bibitem{BellEPRViolation}
J.~S. Bell, \enquote{On the Problem of Hidden Variables in Quantum Mechanics},
  \href{https://dx.doi.org/10.1103/RevModPhys.38.447}{\emph{Rev. Mod. Phys.}
  \textbf{38} (1966) 447}.

\bibitem{EPRBExperiment}
M.~Czachor, \enquote{Einstein-Podolsky-Rosen-Bohm experiment with relativistic
  massive particles},
  \href{https://dx.doi.org/10.1103/PhysRevA.55.72}{\emph{Phys. Rev. A}
  \textbf{55} (1997) 72}.

\bibitem{GingrichAdami}
R.~M. Gingrich and C.~Adami, \enquote{Quantum Entanglement of Moving Bodies},
  \href{https://dx.doi.org/10.1103/PhysRevLett.89.270402}{\emph{Phys. Rev.
  Lett.} \textbf{89} (2002) 270402}.

\bibitem{PeresScudoTerno}
A.~Peres, P.~F. Scudo and D.~R. Terno, \enquote{Quantum Entropy and Special
  Relativity},
  \href{https://dx.doi.org/10.1103/PhysRevLett.88.230402}{\emph{Phys. Rev.
  Lett.} \textbf{88} (2002) 230402}.

\bibitem{JordanShajiSudarshan}
T.~F. Jordan, A.~Shaji and E.~C.~G. Sudarshan, \enquote{Lorentz transformations
  that entangle spins and entangle momenta},
  \href{https://dx.doi.org/10.1103/physreva.75.022101}{\emph{Physical Review A}
  \textbf{75[2]}}.

\bibitem{AhnLeeMoonHwang}
D.~Ahn, H.-j. Lee, Y.~H. Moon and S.~W. Hwang, \enquote{Relativistic
  entanglement and Bell's inequality},
  \href{https://dx.doi.org/10.1103/PhysRevA.67.012103}{\emph{Phys. Rev. A}
  \textbf{67} (2003) 012103}.

\bibitem{AhnLeeHwang}
D.~Ahn, H.-j. Lee and S.~W. Hwang, \enquote{Lorentz-covariant
  reduced-density-operator theory for relativistic-quantum-information
  processing}, \href{https://dx.doi.org/10.1103/PhysRevA.67.032309}{\emph{Phys.
  Rev. A} \textbf{67} (2003) 032309}.

\bibitem{AlsingMilburn}
P.~M. Alsing and G.~J. Milburn, \enquote{Teleportation with a Uniformly
  Accelerated Partner},
  \href{https://dx.doi.org/10.1103/PhysRevLett.91.180404}{\emph{Phys. Rev.
  Lett.} \textbf{91} (2003) 180404}.

\bibitem{LeeYoung}
D.~Lee and E.~Chang-Young, \enquote{Quantum entanglement under Lorentz boost},
  \href{https://dx.doi.org/10.1088/1367-2630/6/1/067}{\emph{New Journal of
  Physics} \textbf{6[1]} (2004) 67}.

\bibitem{Chakrabarti_2009}
A.~Chakrabarti, \enquote{Entangled states, Lorentz transformations and spin
  precession in magnetic fields},
  \href{https://dx.doi.org/10.1088/1751-8113/42/24/245205}{\emph{Journal of
  Physics A: Mathematical and Theoretical} \textbf{42[24]} (2009) 245205}.

\bibitem{Czachor}
M.~Czachor, \enquote{Comment on ``Quantum Entropy and Special Relativity''},
  \href{https://dx.doi.org/10.1103/PhysRevLett.94.078901}{\emph{Phys. Rev.
  Lett.} \textbf{94} (2005) 078901}.

\bibitem{CzachorWilczewski}
M.~Czachor and M.~Wilczewski, \enquote{Relativistic Bennett-Brassard
  cryptographic scheme, relativistic errors, and how to correct them},
  \href{https://dx.doi.org/10.1103/PhysRevA.68.010302}{\emph{Phys. Rev. A}
  \textbf{68} (2003) 010302}.

\bibitem{FuentesMann}
I.~Fuentes-Schuller and R.~B. Mann, \enquote{Alice Falls into a Black Hole:
  Entanglement in Noninertial Frames},
  \href{https://dx.doi.org/10.1103/PhysRevLett.95.120404}{\emph{Phys. Rev.
  Lett.} \textbf{95} (2005) 120404}.

\bibitem{AlsingFuentesMannTessier}
P.~M. Alsing, I.~Fuentes-Schuller, R.~B. Mann and T.~E. Tessier,
  \enquote{Entanglement of Dirac fields in noninertial frames},
  \href{https://dx.doi.org/10.1103/PhysRevA.74.032326}{\emph{Phys. Rev. A}
  \textbf{74} (2006) 032326}.

\bibitem{CabanRembielinski}
P.~Caban and J.~Rembieli\ifmmode~\acute{n}\else \'{n}\fi{}ski,
  \enquote{Lorentz-covariant reduced spin density matrix and
  Einstein-Podolsky-Rosen--Bohm correlations},
  \href{https://dx.doi.org/10.1103/PhysRevA.72.012103}{\emph{Phys. Rev. A}
  \textbf{72} (2005) 012103}.

\bibitem{CabanRembielinski2}
P.~Caban and J.~Rembieli\ifmmode~\acute{n}\else \'{n}\fi{}ski,
  \enquote{Einstein-Podolsky-Rosen correlations of Dirac particles: Quantum
  field theory approach},
  \href{https://dx.doi.org/10.1103/PhysRevA.74.042103}{\emph{Phys. Rev. A}
  \textbf{74} (2006) 042103}.

\bibitem{JordanShajiSudarshan2}
T.~F. Jordan, A.~Shaji and E.~C.~G. Sudarshan, \enquote{Maps for Lorentz
  transformations of spin},
  \href{https://dx.doi.org/10.1103/PhysRevA.73.032104}{\emph{Phys. Rev. A}
  \textbf{73} (2006) 032104}.

\bibitem{LamataMartinSolano}
L.~Lamata, M.~A. Martin-Delgado and E.~Solano, \enquote{Relativity and Lorentz
  Invariance of Entanglement Distillability},
  \href{https://dx.doi.org/10.1103/PhysRevLett.97.250502}{\emph{Phys. Rev.
  Lett.} \textbf{97} (2006) 250502}.

\bibitem{MoonHwangAhn}
Y.~H. Moon, S.~W. Hwang and D.~Ahn, \enquote{Relativistic Entanglements of Spin
  1/2 Particles with General Momentum},
  \href{https://dx.doi.org/10.1143/PTP.112.219}{\emph{Progress of Theoretical
  Physics} \textbf{112[2]} (2004) 219}.

\bibitem{PeresTerno}
A.~Peres and D.~R. Terno, \enquote{Quantum information and relativity theory},
  \href{https://dx.doi.org/10.1103/RevModPhys.76.93}{\emph{Rev. Mod. Phys.}
  \textbf{76} (2004) 93}.

\bibitem{AdessoFuentesEricsson}
G.~Adesso, I.~Fuentes-Schuller and M.~Ericsson, \enquote{Continuous-variable
  entanglement sharing in noninertial frames},
  \href{https://dx.doi.org/10.1103/PhysRevA.76.062112}{\emph{Phys. Rev. A}
  \textbf{76} (2007) 062112}.

\bibitem{BrandaoGour}
F.~G. S.~L. Brand\~ao and G.~Gour, \enquote{Reversible Framework for Quantum
  Resource Theories},
  \href{https://dx.doi.org/10.1103/PhysRevLett.115.070503}{\emph{Phys. Rev.
  Lett.} \textbf{115} (2015) 070503}.

\bibitem{HorodeckiOppenHeim}
M.~HORODECKI and J.~OPPENHEIM, \enquote{(QUANTUMNESS IN THE CONTEXT OF)
  RESOURCE THEORIES},
  \href{https://dx.doi.org/10.1142/S0217979213450197}{\emph{International
  Journal of Modern Physics B} \textbf{27[01n03]} (2013) 1345019},
  \href{https://arxiv.org/abs/https://doi.org/10.1142/S0217979213450197}{{\tt
  arXiv:https://doi.org/10.1142/S0217979213450197}}.

\bibitem{GourSpekkens}
G.~Gour and R.~W. Spekkens, \enquote{The resource theory of quantum reference
  frames: manipulations and monotones},
  \href{https://dx.doi.org/10.1088/1367-2630/10/3/033023}{\emph{New Journal of
  Physics} \textbf{10[3]} (2008) 033023}.

\bibitem{GourMarvianSpekkens}
G.~Gour, I.~Marvian and R.~W. Spekkens, \enquote{Measuring the quality of a
  quantum reference frame: The relative entropy of frameness},
  \href{https://dx.doi.org/10.1103/PhysRevA.80.012307}{\emph{Phys. Rev. A}
  \textbf{80} (2009) 012307}.

\bibitem{MarvianSpekkens}
I.~Marvian and R.~W. Spekkens, \enquote{The theory of manipulations of pure
  state asymmetry: I. Basic tools, equivalence classes and single copy
  transformations},
  \href{https://dx.doi.org/10.1088/1367-2630/15/3/033001}{\emph{New Journal of
  Physics} \textbf{15[3]} (2013) 033001}.

\bibitem{MarvianSpekkens2}
I.~Marvian and R.~W. Spekkens, \enquote{Modes of asymmetry: The application of
  harmonic analysis to symmetric quantum dynamics and quantum reference
  frames}, \href{https://dx.doi.org/10.1103/PhysRevA.90.062110}{\emph{Phys.
  Rev. A} \textbf{90} (2014) 062110}.

\bibitem{MarvianSpekkens3}
I.~Marvian and R.~W. Spekkens, \enquote{Extending Noether’s theorem by
  quantifying the asymmetry of quantum states},
  \href{https://dx.doi.org/10.1038/ncomms4821}{\emph{Nat. Commun.} \textbf{5}
  (2014) 3821}.

\bibitem{MarvianSpekkensZanardi}
I.~Marvian, R.~W. Spekkens and P.~Zanardi, \enquote{Quantum speed limits,
  coherence, and asymmetry},
  \href{https://dx.doi.org/10.1103/PhysRevA.93.052331}{\emph{Phys. Rev. A}
  \textbf{93} (2016) 052331}.

\bibitem{MarvianSpekkens4}
I.~Marvian and R.~W. Spekkens, \enquote{How to quantify coherence:
  Distinguishing speakable and unspeakable notions},
  \href{https://dx.doi.org/10.1103/PhysRevA.94.052324}{\emph{Phys. Rev. A}
  \textbf{94} (2016) 052324}.

\bibitem{ChitambarGour}
E.~Chitambar and G.~Gour, \enquote{Comparison of incoherent operations and
  measures of coherence},
  \href{https://dx.doi.org/10.1103/PhysRevA.94.052336}{\emph{Phys. Rev. A}
  \textbf{94} (2016) 052336}.

\bibitem{YaoDongXiaoSun}
Y.~Yao, G.~H. Dong, X.~Xing and C.~P. Sun, \enquote{Frobenius-norm-based
  measures of quantum coherence and asymmetry},
  \href{https://dx.doi.org/10.1038/srep32010}{\emph{Sci. Rep.} \textbf{6}
  (2016) 32010}.

\bibitem{WangFangYu}
W.-C. Wang, M.-F. Fang and M.~Yu, \enquote{Intrinsic basis-independent quantum
  coherence measure},  2017.

\bibitem{MSG2024quantum}
A.~Mukherjee, S.~Sen and S.~Gangopadhyay, \enquote{Quantum coherence measures
  for generalized Gaussian wave packets under a Lorentz boost},
  \href{https://dx.doi.org/10.1103/PhysRevA.110.052413}{\emph{Phys. Rev. A}
  \textbf{110} (2024) 052413}.

\bibitem{chatterjee2017preservation}
R.~Chatterjee and A.~S. Majumdar, \enquote{Preservation of quantum coherence
  under Lorentz boost for narrow uncertainty wave packets},
  \href{https://dx.doi.org/10.1103/PhysRevA.96.052301}{\emph{Phys. Rev. A}
  \textbf{96} (2017) 052301}.

\bibitem{FriisBertlmannHuber}
N.~Friis, R.~A. Bertlmann, M.~Huber and B.~C. Hiesmayr, \enquote{Relativistic
  entanglement of two massive particles},
  \href{https://dx.doi.org/10.1103/PhysRevA.81.042114}{\emph{Phys. Rev. A}
  \textbf{81} (2010) 042114}.

\bibitem{weinberg1995quantum}
S.~Weinberg, The Quantum Theory of Fields: Volume I: Foundations, Cambridge
  Monographs on Mathematical Physics, Cambridge University Press, Cambridge, UK
  1995.
  
\bibitem{l1_Norm}
T.~Baumgratz, R.~M. Cramer, and M.~B.~Pleino, \enquote{Quantifying Coherence},
  \href{https://link.aps.org/doi/10.1103/PhysRevLett.113.140401}{\emph{Phys. Rev. Lett}
  \textbf{113} (2014) 140401}.
  


\end{thebibliography}
\end{document}